\documentclass[USenglish,twocolumn]{article}

\ifx\directlua\undefined\ifx\XeTeXcharclass\undefined
  \usepackage[utf8]{inputenc}                           
  \else\RequirePackage[no-math]{fontspec}[2017/03/31]\fi 
  \else\RequirePackage[no-math]{fontspec}[2017/03/31]\fi 
\usepackage[sort&compress,square,numbers]{natbib}
\usepackage[big,online]{dgruyter}
\usepackage{comment}

\theoremstyle{dgthm}

\theoremstyle{dgdef}

\begin{document}

\articletype{Research Article}

\author[1]{Timon Eichhorn}
\author[2]{Nicholas Jobbitt}
\author[2]{Sören Bieling}
\author[3]{Shuping Liu}
\author[7]{Tobias Krom}
\author[4]{Diana Serrano}
\author[5]{Robert Huber}
\author[6]{Ulrich Lemmer}
\author[8]{Hugues de Riedmatten}
\author[4]{Philippe Goldner}
\author*[9]{David Hunger} 
\affil[1]{Physikalisches Institut, Karlsruhe Institute of Technology (KIT), Karlsruhe, Germany; and Institute for Quantum Materials and Technologies (IQMT), Karlsruhe Institute of Technology (KIT), Eggenstein-Leopoldshafen, Germany, timon.eichhorn@kit.edu; https://orcid.org/0000-0003-3469-8876}
\affil[2]{Physikalisches Institut, Karlsruhe Institute of Technology (KIT), Karlsruhe, Germany}
\affil[3]{Shenzhen Institute for Quantum Science and Engineering, Southern University of Science and Technology, Shenzhen, China}
\affil[4]{Chimie ParisTech, PSL University, CNRS, Institut de Recherche de Chimie Paris, Paris, France}
\affil[5]{Lichttechnisches Institut, Karlsruhe Institute of Technology (KIT), Karlsruhe, Germany; https://orcid.org/0000-0002-7457-1688}
\affil[6]{Lichttechnisches Institut, Karlsruhe Institute of Technology (KIT), Karlsruhe, Germany}
\affil[7]{Physikalisches Institut, Universität Heidelberg, Heidelberg, Germany}
\affil[8]{ICFO-Institut de Ciencies Fotoniques, The Barcelona Institute of Science and Technology, Barcelona, Spain}
\affil[9]{Physikalisches Institut, Karlsruhe Institute of Technology (KIT), Karlsruhe, Germany; and Institute for Quantum Materials and Technologies (IQMT), Karlsruhe Institute of Technology (KIT), Eggenstein-Leopoldshafen, Germany, david.hunger@kit.edu}

\title{Multimodal Purcell enhancement and optical coherence of $\text{Eu}^{\text{3+}}$ ions in a single nanoparticle coupled to a microcavity}
\runningtitle{Multimodal Purcell-enhancement}
\abstract{Europium-doped nanocrystals constitute a promising material for a scalable future quantum computing platform. Long-lived nuclear spin states could serve as qubits addressed via coherent optical transitions. In order to realize an efficient spin-photon interface, we couple the emission from a single nanoparticle to a fiber-based microcavity under cryogenic conditions. The spatial and spectral tunability of the cavity permits us to place individual nanoparticles in the cavity, to measure the inhomogeneous linewidth of the ions, and to show a multi-modal Purcell-enhancement of two transition in Eu$^{\text{3+}}$. A halving of the free-space lifetime to 1.0\,ms is observed, corresponding to a 140-fold enhancement of the respective transition. Furthermore, we observe a narrow optical linewidth of 3.3\,MHz for a few-ion ensemble in the center of the inhomogeneous line. The results represent an important step towards the efficient readout of single Eu$^{\text{3+}}$ ions, a key requirement for the realization of single-ion-level quantum processing nodes in the solid state.}
\journalname{Nanophotonics}
\journalyear{2024}
\journalvolume{aop}

\maketitle

\section{Introduction}
Rare-earth ions (REI) doped into inorganic crystals constitute a promising platform for future quantum technologies due to their long optical and spin coherence times. Among the various species of REIs, europium features among the narrowest optical linewidths (sub-kHz) \cite{macfarlane_sub-kilohertz_1981} and longest spin coherence lifetimes of over 6\,h \cite{zhong_optically_2015}, making it an ideal candidate for storing quantum information. Furthermore, state-selective optical transitions connected to the hyperfine (qubit) states open up the possibility for fast, all-optical single-qubit gate operations \cite{kinos_designing_2021}. The difference in the permanent electric dipole moments of the ground and optically excited states gives rise to a dipole blockade, similar to the Rydberg blockade \cite{jaksch_fast_2000}, which can be used to entangle two qubits \cite{ohlsson_quantum_2002} at few-nanometer distances. The beneficial ratio between the inhomogeneous and the narrow homogeneous optical linewidth of about $10^3-10^5$ opens up a way to spectrally address many individual ions. To meet these requirements, a limited number of ions has to be spatially confined, e.g. in a nanoparticle. Such a nanoscale quantum processor, based on europium ions, would then be a scalable alternative to existing quantum processor architectures with the potential for high qubit interconnectivity \cite{kinos_high-connectivity_2022} and high gate fidelities \cite{kinos_designing_2021,kinos_roadmap_2021}.

The dipole-forbidden nature of the coherent 4f-4f transitions in REIs, which are weakly induced in low-symmetry crystals, leads to long lifetimes ($\sim$ms) of the optically excited state. This renders single ion readout difficult due to low fluorescence count rates. However, enhancing the density of states of electromagnetic modes and increasing their vacuum field by an optical cavity gives rise to the Purcell effect \cite{purcell_spontaneous_1946}, which manifests as a lifetime shortening. The Purcell factor $F_{\text{P}}$, quantifying the increase in the spontaneous emission rate, for a Fabry-Pérot type cavity is given as:
\begin{equation}
    F_{\text{P}} = \frac{6}{\pi^3}\,\left(\frac{\lambda}{n}\right)^2\,\frac{\mathcal{F}}{w_0^2}.
    \label{eq:Purcell-factor}
\end{equation}
Here, $\mathcal{F}$ denotes the cavity finesse, $w_0$ is the mode waist of the Gaussian cavity mode, $\lambda$ the transition wavelength, and $n$ the refractive index. Using nano- and micro-scale optical cavities, single REI readout has been demonstrated for various species in the recent years, but not for Eu$^\text{3+}$ to date \cite{dibos_atomic_2018,kindem_control_2020,chen_parallel_2020,xia_tunable_2022,ulanowski_spectral_2022,deshmukh_detection_2023}.

\begin{figure}
\includegraphics[width=\columnwidth]{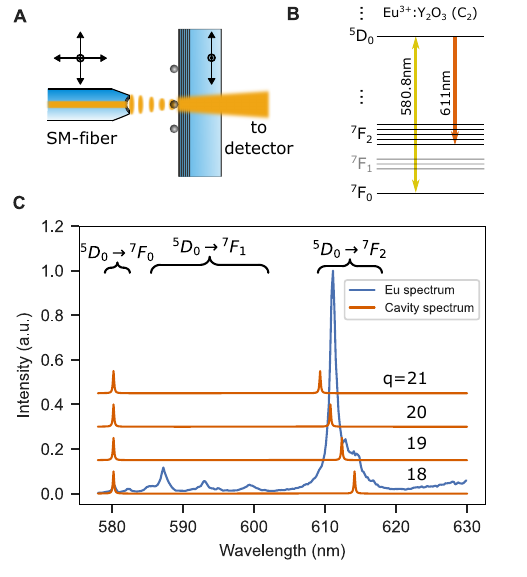}
\caption[Fiber Fabry-Pérot setup and europium level scheme]{\textbf{A}: schematic drawing of the fiber-based Fabry-Pérot microcavity. Laser light enters the cavity via a single mode optical fiber (SM-fiber), and the transmission is collected by a photodiode or a single photon counting module (see optical setup in supplementary material). \textbf{B}: Relevant level scheme of europium. \textbf{C}: emission spectrum of the $^5D_0$ excited state for the transitions depicted in B (blue line), together with a simulation of the cavity spectrum for longitudinal mode orders $q=18-21$ of the 580\,nm resonance (red lines).}
\label{fig:FFPC+level-scheme}
\end{figure}

Here, we present important steps towards the detection of single europium ions. We study the optical and coherence properties of small ion ensembles inside individual europium-doped nanoparticles (NPs) at cryogenic temperatures. To reach a high Purcell factor, we incorporate the NPs into a fiber-based Fabry-Pérot microcavity \cite{hunger_fiber_2010,pfeifer_achievements_2022}, which features a high finesse and a small mode waist. The cavity's spatial tunability allows us to study six different NPs and to perform cavity-enhanced spectroscopy. The spectral tunability of the cavity is then employed to demonstrate a multimodal Purcell-enhancement of two decay channels of the same excited state. Furthermore, we use transient spectral hole burning to determine an upper bound of the homogeneous linewidth. The measured parameters allow for a realistic estimation of the count rate that we expect from a single europium ion. The system performance demonstrated here should enable the detection and characterization of single ions.

\section{Methods}
We operate a fully-tunable, open-access, fiber-based Fabry-Pérot microcavity inside a liquid helium flow cryostat. The cavity assembly is mounted on the cold plate of the cryostat and is cooled by helium exchange gas down to a minimum temperature of 3.5\,K. The microcavity consists of a concave fiber mirror and a macroscopic planar mirror carrying the nanoparticles as shown in Fig.\,\ref{fig:FFPC+level-scheme}A. We use a single-mode (SM) optical fiber where a concave profile with radius of curvature of 25\,µm has been machined onto the end facet by $\text{CO}_{\text{2}}$ laser machining \cite{hunger_laser_2012,muller_ultrahigh-finesse_2010}. The fiber is then coated with a distributed Bragg reflector (DBR) mirror with 25\,ppm transmission. The planar opponent is a DBR mirror with higher transmission of 200\,ppm to optimize the out-coupled rate of photons $R_\text{out}$ (see below) in the presence of mirror loss and nanoparticle scattering. Together with the absorption and scattering losses of the mirrors, this results in a bare cavity finesse of 17,500 at 580\,nm and 9,500 at 611\,nm. The mode waist at the planar mirror is about 1.4\,µm, resulting in a nominal Purcell factor of $F_{\text{P}}=$580 (330) at 580\,nm (611\,nm) according to Eq.\,\ref{eq:Purcell-factor}. The fiber position can be fine tuned with sub-picometer precision in all directions by piezo actuators. Electrical motors achieve a coarse tuning of the cavity length over hundreds of microns and lateral movement of the mirror of up to one mm \cite{pallmann_highly_2023}. The nanopositioning stage is fully operational in the temperature range from room temperature down 3.5\,K. The root mean square cavity length jitter, which has a strong influence on the Purcell-effect, can be as low as 2.5\,pm under active stabilization \cite{pallmann_highly_2023}. For the measurements presented here, the cavity length jitter was around 8\,pm under active stabilization for all measurements except the pulsed lifetime histograms. For those, we only applied a slow drift compensation to stay on resonance resulting in a few picometer worse cavity stability.

The investigated sample are Eu$^{3+}$:Y$_2$O$_3$ nanoparticles (NP) of an average diameter of 60\,nm, which are doped with europium ions at a concentration of 0.3\,\%. The NPs were synthesized by homogeneous precipitation followed by an annealing step of 18\,h at 800°C and a high-power oxygen plasma annealing ($2\times3$\,min at 900\,W power using a 2.45\,GHz microwave source) \cite{bartholomew_optical_2017,liu_controlled_2018,liu_defect_2020,fossati_optical_2023}. The small particle size was chosen to achieve small scattering losses inside the cavity, which amount to about 13\,ppm roundtrip losses for a 60\,nm particle.

The relevant level structure of Eu$^{3+}$ at the $\text{C}_{\text{2}}$ point-symmetry site of yttria is sketched in Fig.\,\ref{fig:FFPC+level-scheme}B. We are exciting the ions via the ground state transition from the $^7F_0$ to the $^5D_0$ level at 580.8\,nm with a frequency doubled diode laser (\textit{Toptica DLPro}) at cryogenic temperatures. This transition shows a narrow homogenous linewidth of down to 116\,kHz as measured by photon echoes in a powder of NPs at 1.4\,K (see Fig.\,6 in supplementary material). However, the $^5D_0 \rightarrow\, ^7F_0$ transition in Eu$^{3+}$:Y$_2$O$_3$ shows a low branching ratio, and we determine this quantity by recording a broadband spectrum which includes all transitions, and compare the $^5D_0\rightarrow\,^7F_0$ emission with the integrated emission. This yields $\zeta_{\text{580nm}}=0.7(1)$\,\%, resulting in a strongly reduced coupling to the cavity. Therefore, we also study cavity coupling of the $^5D_0\rightarrow\,^7F_2$ transition at 611\,nm which has a higher branching ratio of 36(3)\% but features a broad homogeneous linewidth of 680(20)\,GHz at 4.2\,K due to phonon relaxation, see Fig.\,\ref{fig:FFPC+level-scheme}C.
The tunability of the cavity length $d_c$ enables us to control the free-spectral range (FSR) of the cavity, $\text{FSR} = c/(2\,d_c)$. This can be used to spectrally overlap two consecutive resonances of the cavity with the 580.8\,nm and 611\,nm transitions of europium, see Fig.\,\ref{fig:FFPC+level-scheme}C. Tuning the cavity to the 20th longitudinal mode order $q$ at 580.8\,nm yields the double resonance condition.

\begin{figure}
\includegraphics[width=\columnwidth]{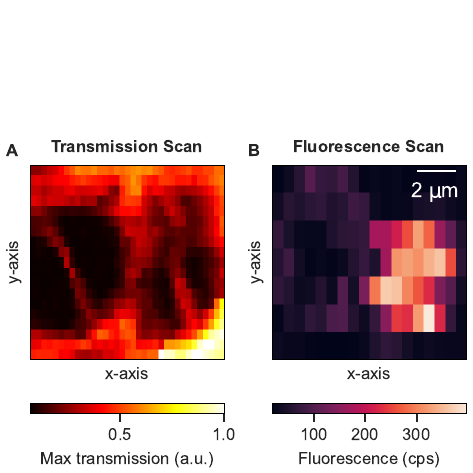}
\caption[Cavity transmission and fluorescence scans]{Scanning cavity microscopy scans recording the peak transmission through the cavity (\textbf{A}) and the fluorescence count rate (\textbf{B}) of the same region on the planar mirror.}
\label{fig:cavity_scans}
\end{figure}

To distribute individual NPs homogeneously on the planar mirror, we examined a novel method using an aerosol printer (\textit{Aerosol Jet 5X, Optomec Inc.}). Therefore, the powder of NPs is dispersed in water together with the surfactant sodium dodecyl sulfate (SDS) to form a colloidal solution. The latter is used as an ink in the aerosol printer, which focuses an aerosol stream with droplet sizes of a few microns onto the mirror surface \cite{gramlich_process_2023,huber_gedruckte_2024}. By carefully adjusting the different printer parameters, we obtain a homogeneous distribution of single nanoparticles as well as small agglomerations with an average inter-particle distance larger than the cavity mode diameter of about 3\,µm. A certain nanoparticle can then be located inside the cavity by imaging the scattering losses using scanning cavity microscopy \cite{mader_scanning_2015}. A transmission scan of two large particles or agglomerations can be seen in Fig.\,\ref{fig:cavity_scans}A. Here, one can observe the TEM$_{01}$ cavity mode profile, yielding the dominant signal contribution in this particular measurement. For comparison, a scanning cavity image of different NPs in Section\,2 of the supplementary material shows the TEM$_{00}$ as well as TEM$_{10}$ modes of the cavity within one scan. Since not every scatterer on the mirror turns out to be an europium-doped NP, we additionally employ a fluorescence cavity scan method. In Fig.\,\ref{fig:cavity_scans}B the same area is scanned as in Fig.\,\ref{fig:cavity_scans}A, and the cavity-enhanced fluorescence of the 611\,nm transition is recorded while the 580\,nm laser continuously excites the ions via the double resonance. Simultaneously, the laser transmission is used to actively stabilize the cavity to the side-of-fringe. It can be seen that only the right scatterer shows an increased fluorescence count rate of more than 300 counts per second (cps) and thus contains europium ions. The fluorescence distribution reveals the dominantly coupling cavity mode, which is the fundamental TEM$_{00}$ mode.

\section{Inhomogeneous linewidths}
\begin{figure}
\includegraphics[width=\columnwidth]{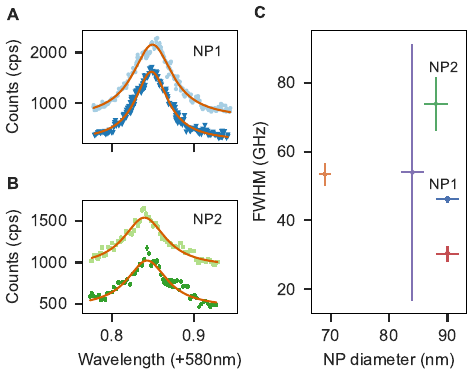}
\caption[Inhomogeneous linewidths of different nanoparticles]{PLE scans of the inhomogeneous line of europium ions from two different nanoparticles (\textbf{A} and \textbf{B}) at 20\,K. A Lorentzian line (red) is fit to the data to extract the full width at half maximum (FWHM). A second scan (light blue and light green) is offset vertically to show the reproducibility of the measurement. In total, five different nanoparticles were measured in this way. \textbf{C} The FWHMs are plotted against the corresponding diameter of the nanoparticle.}
\label{fig:inhomo-lines}
\end{figure}

After identifying suitable NPs as described above, we  perform cavity-enhanced photoluminescence excitation (PLE) spectroscopy of the inhomogeneous line of the $^7F_0\rightarrow\,^5D_0$ transition. Therefore, we stabilize the cavity on the 20th longitudinal mode order at 580\,nm and continuously excite the ions under strong power broadening at intracavity laser powers of several hundred microwatts. We choose a low cavity locking setpoint in order to achieve a stable intracavity power during the frequency scan. Here, we use a dye laser (\textit{Matisse 2 DX, Sirah Lasertechnik}) which enables mode-hop-free scans over 60\,GHz at a speed of 500\,MHz/s, while the cavity-coupled 611\,nm fluorescence light is recorded by a single photon counting module (\textit{COUNT-100C, LaserComponents}) with a dark count rate of 20\,Hz. Scans over the inhomogeneous line of two different NPs are shown in Fig.\,\ref{fig:inhomo-lines}A and B. To cover the full spectrum, three scans are stitched together. Each plot contains two measurements under identical conditions (offset for better visibility), proving the reproducibility of the measurement. Slight deviations from the Lorentzian fit (red line displays fit) are visible for both particles which reproduce in each scan. We attribute this to the statistical fine structure, i.e. the Poissonian distribution of the ion number within a power-broadened frequency interval. Since the total number of europium ions inside a single nanoparticle is on the order of $10^3-10^4$, we expect a significant fluctuation of the number of ions per frequency interval, which manifests in a fluctuation of the count rate dependent on the laser frequency.

Plot C in Fig.\,\ref{fig:inhomo-lines} summarizes the inhomogeneous linewidths for five different nanoparticles that were investigated. The diameter of each particle can be determined with a high precision from the observed scattering losses, $S_{\text{NP}}\propto d_{\text{NP}}^6$, on a scanning cavity image as depicted in Fig.\,\ref{fig:cavity_scans}A. Overall, we find inhomogeneous linewidths between 30\,GHz and 74\,GHz, without any correlation to the NP diameter. A reference measurement on a macroscopic amount of powder of the same NP batch yielded a value of 34\,GHz, see Fig.\,3 in the supplementary information.  These linewidths are broader than the 10\,GHz inhomogeneous broadening found in yttria bulk crystals \cite{macfarlane_sub-kilohertz_1981} and the 22\,GHz found in \cite{casabone_cavity-enhanced_2018} on a yttria NP of slightly larger diameter. We attribute this broadening to the oxygen plasma annealing step in the NP fabrication as reported in \cite{liu_defect_2020,fossati_optical_2023}, which was found to improve the homogeneous linewidth. The larger inhomogeneous linewidth compared to bulk crystals could be beneficial for single ion addressing and spectral multiplexing due to a lower spectral ion density.

\section{Multimodal Purcell enhancement}
\begin{figure}
\includegraphics[width=\columnwidth]{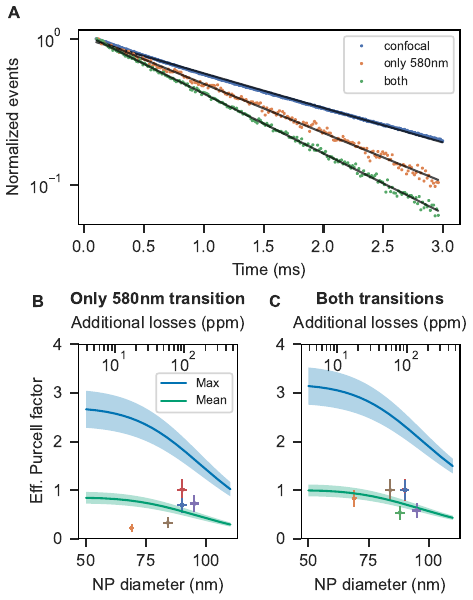}
\caption[Lifetime histogram and effective Purcell factors]{\textbf{A}: lifetime histograms of the free-space emission in a confocal microscope (blue) and inside the cavity (orange and green). Tuning the cavity length is used to Purcell-enhance only the 580\,nm transition (orange) or both the 580\,nm and 611\,nm transitions simultaneously (green). \textbf{B} and \textbf{C}: comparison of the effective Purcell factors measured for six different nanoparticles together with a theoretical estimation of the maximum (blue) and ensemble averaged (green) Purcell factor dependent on the nanoparticle size. In \textbf{B,} only the 580\,nm transition is enhanced by the cavity, whereas \textbf{C} shows the multimodal Purcell-enhancement when the cavity is resonant with both transitions.}
\label{fig:Purcell+Lifetime-comparison}
\end{figure}

Next, we quantify the lifetime shortening of the europium ions due to the Purcell effect in the cavity. The effective Purcell factor can be determined experimentally by comparing the excited state lifetime in the cavity $T_{1,\text{c}}$ with the free-space lifetime $T_1$, i.e. without cavity enhancement:
\begin{equation*}
    F_{\text{P}}^{\text{eff}} = \zeta\,F_{\text{P}} = \frac{T_1}{T_{1,\text{c}}} - 1.
    \label{eq:Eff_Purcell-factor}
\end{equation*}
To measure the free-space lifetime, we recorded a photon count histogram under pulsed excitation with a confocal microscope at room temperature. Here, we excite all ions within a NP by off-resonant excitation using a 532\,nm laser (\textit{Cobolt Samba, Hübner Photonics}) and collecting the fluorescence of the 611\,nm transition. An exemplary histogram of the exponential decay is shown in Fig.\,\ref{fig:Purcell+Lifetime-comparison} A together with a fit (grey line). The average determined lifetime of four different nanoparticles residing on the cavity mirror gives a value of $T_1=2.0\pm 0.1$\,ms. For some nanoparticles, we also measured the free-space lifetime at cryogenic temperatures inside the cavity under single resonance conditions by detuning the cavity resonance from the laser by six \,HWHM of the cavity linewidth. This permits us to resonantly excite and detect a sub-ensemble of ions but suppressing the Purcell enhancement to about 3\% of the Purcell factor on resonance. A negligible overlap between the cavity and free-space modes avoids Purcell suppression of the fluorescence. Such measurements yield consistent results with the room temperature confocal measurements. To retrieve the cavity-enhanced lifetime, we make use of the cavity length tunability and selectively enhance only the 580\,nm transition, only the 611\,nm, or both transitions simultaneously. Since there is no spectral overlap between both transitions, the Purcell factors add up linearly when both transitions are resonant with the cavity (see section 6 in supplementary material):
\begin{equation*}
    F_{\text{P,both}} = F_{\text{P,580nm}} + F_{\text{P,611nm}}.
    \label{eq:Purcell-factor_sum}
\end{equation*}
Therefore, we expect and observe the shortest lifetimes for the multi-modal Purcell enhancement as shown by the green datapoints in Fig.\,\ref{fig:Purcell+Lifetime-comparison}A. For this NP, we measure a lifetime of 1.3(0.1)\,ms for the 580\,nm transition and 1.1(0.1)\,ms for the twofold Purcell enhancement.

The resulting effective Purcell factors that we measured for six different nanoparticles are summarized in Fig.\,\ref{fig:Purcell+Lifetime-comparison}B(C) for enhancing only the 580\,nm (both) transition(s). The solid green (blue) line shows the calculation of the ensemble averaged (maximum) effective Purcell factors dependent on the size of the NP together with the error interval of plus and minus one standard deviation (shaded regions). The ensemble averaged Purcell factor takes into account the random dipole orientations of the emitters due to the polycrystalline nature of the nanoparticles. Furthermore, an average of the electric field amplitude at the random positions of the ions inside the nanoparticle is taken into account. The maximum effective Purcell factor assumes a perfect overlap between the dipole and electric field vectors and a position at the field maximum of the standing wave. In both cases the RMS cavity length jitter of 8\,pm was taken into account using the formula given in \cite{pallmann_highly_2023}. Details on the derivation of the effective Purcell factor can be found in the supplementary material.

We measured a mean, ensemble-averaged, effective Purcell factor of 0.6(0.3) for enhancing only the 580\,nm transition and a value of 0.8(0.2) for both transitions. In both cases, the maximal observed effective Purcell factor is 1.0(0.2). This agrees well with the theoretical estimation for the sizes of the particular NPs. Therefore, we can deduce a Purcell-enhancement of up to 2.5 and 3.0 for a perfectly coupling single europium ion inside a 60\,nm NP. Although, we only observe a halving of the free-space lifetime inside the cavity, this amounts to a Purcell factor of $F_{\text{P}}=F_{\text{P}}^{\text{eff}} / \zeta_{\text{580nm}}=140$ and an increase of the branching ratio to $\zeta_{\text{c}}=(F_{\text{P}}^{\text{eff}}+\zeta_{\text{580nm}})/(F_{\text{P}}^{\text{eff}}+1)=0.5$.

The two transitions couple very differently to the cavity: The 580\,nm transition has a low branching ratio $\zeta=0.7\%$ and is significantly narrower than the cavity linewidth $\kappa$ and thus couples in the bad cavity regime. On the other hand, the 611\,nm transition has a dominant branching ratio $\zeta=36\%$ and features a broad homogeneous linewidth of $\Gamma_{\text{h}}=680(20)$\,GHz, which leads to the bad emitter regime. In this regime the effective Purcell factor reduces according to $F_{\text{P}}^{\text{eff}}\propto \kappa / \Gamma_{\text{h}}=0.004$. This results in an ensemble-averaged (maximal) effective Purcell factor of 0.15 (0.47) for enhancing solely the 611\,nm transition. Still, enhancing the 611\,nm transition additionally to the 580\,nm transition leads to a total effective Purcell factor of $F_{P,\text{both}}=2.5+0.47=2.97$ for a perfectly coupling ion. Thus, making use of the multimodal Purcell enhancement and coupling also the 611\,nm transition to the cavity, increases the Purcell factor by about $0.47/2.5\approx 20\%$.

\section{Optical coherence}
\begin{figure}
\includegraphics[width=\columnwidth]{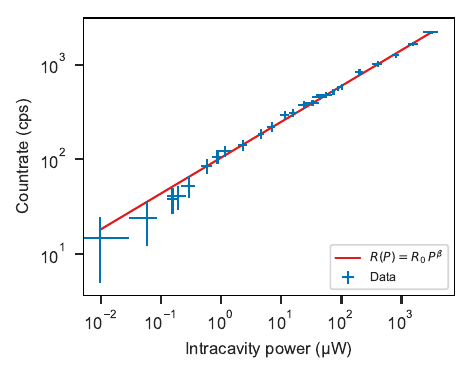}
\caption[Saturation curve of a nanoparticle at 20\,K]{Saturation behaviour in the center of the inhomogeneous line at 20\,K. The data is background-corrected and a fit of a power law (red) is shown.}
\label{fig:saturation}
\end{figure}

Finally, we determine the homogeneous linewidth of a sub-ensemble of ions in the center of the inhomogeneous line at 4.3\,K using transient spectral hole burning \cite{volker_hole-burning_1989,gritsch_narrow_2022}. This method requires saturation of the transition, and we therefore study the power dependence of the emission rate. Coherently driving the ions with a high laser power leads to power-broadening of the linewidth $\Gamma_h(P)\propto \sqrt{P}$. The high spectral density of emitters in the center of the inhomogeneous line thus gives rise to an increase in the fluorescence count rate according to $R(P)\propto\sqrt{P}$ since more detuned emitters are excited with increased laser power. This can be seen in the measured saturation curve shown in Fig.\,\ref{fig:saturation}. We find a good agreement of the data with the fit of a power law $R(P)=R_0\,P^{\beta}$, resulting in a scaling factor $R_0=105(5)$\,cps and an exponent of $\beta=0.38(0.01)$. The deviation from a square root behaviour is ascribed to different coupling strengths of the ions within the sub-ensemble.

\begin{figure}
\includegraphics[width=\columnwidth]{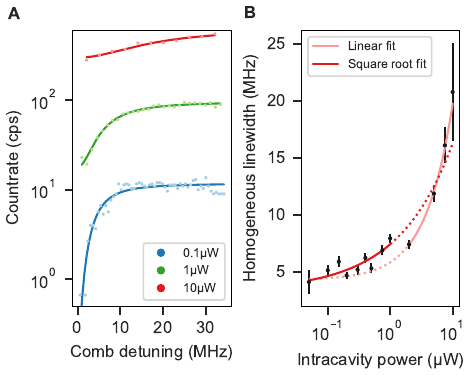}
\caption[Transient spectral hole burning measurement at 4\,K]{\textbf{A}: transient spectral holes at three different intracavity power levels. The solid lines display fits of an inverted Lorentzian line. \textbf{B}: half-width of the transient spectral hole as a function of the intracavity power. Fitting the square root function of Eq.\,\ref{eq:power_broadening_law} (dark red) to the data permits to extrapolate an upper bound of the homogeneous linewidth. The four datapoints at highest power were excluded from the fit since a linear function (light red) yields better agreement.}
\label{fig:TSHB-coherence}
\end{figure}

The homogeneous linewidth at a certain power level can now be determined by imprinting a frequency comb of $N$ teeth onto the laser using an electro-optic modulator and scanning the frequency spacing between the comb teeth. At zero detuning, we expect a count rate of $R(P)=R_0\,\sqrt{N\,P_{\text{tooth}}}$, while if the detuning exceeds the power-broadened linewidth, each tooth excites a separate sub-ensemble of ions and hence the count rate increases by a factor of $\sqrt{N}$ to $R(P)=R_0\,N\,\sqrt{P_{\text{tooth}}}$. The count rate versus detuning follows an inverted Lorentzian line, the transient spectral hole, as can be seen in Fig.\,\ref{fig:TSHB-coherence}A. From a fit, we extract the FWHM of the homogeneous linewidth as $\Gamma_h(P)=0.5\Gamma_{\text{hole}}(P)-\Gamma_{\text{laser}}$ at the respective power and plot these results in Fig.\,\ref{fig:TSHB-coherence}B. The laser linewidth can be neglected, since it is significantly below one Megahertz. The zero-power value $\Gamma_0$ of the fit function,
\begin{equation}
\Gamma_{\text{h}}(P) = \alpha_1 \sqrt{P} + \Gamma_0,
\label{eq:power_broadening_law}
\end{equation}
reveals an upper bound of the homogeneous linewidth, which amounts to 3.3(6)\,MHz, corresponding to a coherence time of $T_2^*=96\pm18$\,ns. The Lorentzian line shape observed in Fig.\,\ref{fig:TSHB-coherence}A and the moderate increase of $\Gamma_{\text{h}}(P)$ compared to $\Gamma_0$ confirm that Eq.\,\ref{eq:power_broadening_law} is a valid model within the power range up to 1\,µW where it is applied. The data points for the four highest powers were excluded from the fit since they match better to a linear function. This suggests a temperature-induced broadening due to laser heating of the nanoparticle, since the homogeneous linewidth shows a linear temperature dependence below 10\,K \cite{bartholomew_optical_2017}.

The measured linewidth is about a factor of two higher than the value of 1.6\,MHz measured at 4\,K by \cite{fossati_optical_2023} using spectral hole burning spectroscopy in a large ensemble, and much larger than the value obtained from photon echo measurements (116\,kHz) as shown in Fig.\,6 in the supplementary information. The comparably broad linewidth can be explained by the high power levels above saturation that are required for this method. In this regime, instantaneous spectral diffusion (ISD) \cite{kinos_microscopic_2022} due to dipole-dipole interactions between resonantly but also off-resonantly excited ions broadens the linewidth. Since all data points remain in this regime, an extrapolation to zero power can not yield a linewidth that is unaffected by ISD as one would expect it for frequency selective single ion addressing.

Using the measured linewidth, $\Gamma_h=2\pi\cdot 3.3\,$MHz, we are able to calculate the saturation intensity $I_{\text{sat}}$ of a perfectly coupling single ion. Starting from the resonant saturation parameter given in \cite{julsgaard_understanding_2007} and expressing the Rabi frequency in terms of cavity QED parameters for a Fabry-Pérot type cavity mode, we arrive at the following expression:
\begin{equation}
    I_{\text{sat}} = \frac{4\pi^3}{3}\hbar c \frac{\Gamma_h}{\zeta \lambda^3}.
    \label{eq:saturation_intensity}
\end{equation}
This results in a value of $I_{\text{sat}}=2\cdot 10^4$\,W/$\text{m}^{\text{2}}$ or a intracavity saturation power of $P_{\text{sat}}=61$\,nW for the 580\,nm transition at 4.3\,K.

We note that with the performed spectroscopy, we already reached the level of the single ion saturation power and probe only a very small number of ions: We simulate the random distribution of the total number of ions (derived from the NP diameter and doping concentration) over the measured inhomogeneous profile taking into account the hyperfine structure to estimate the number of ions that we address within a power-broadened linewidth at the lowest power level of 50\,nW intracavity power in Fig.\,\ref{fig:saturation} and \ref{fig:TSHB-coherence}. This yields about 15(2) europium ions.

\section{Single ion count rate estimation}
The agreement between the measured and estimated ensemble averaged Purcell factors in Fig.\,\ref{fig:Purcell+Lifetime-comparison}B and C as well as the confirmation of a homogeneous linewidth much narrower than the cavity linewidth (bad cavity regime) enables us to estimate the count rate that we expect from a perfectly coupling single europium ion. We simulate the count rate for pulsed, resonant excitation on the 580\,nm transition with a 1\,µs excitation pulse and variable length of the detection time window dependent on the repetition rate $R_{\text{rep}}=1/(t_{\text{ex}}+t_{\text{det}})$. A 3-tone excitation pulse matching the ground state hyperfine splitting is assumed, to avoid optical pumping. Since we expect incoherent excitation, we assume an excited state population of $p_{\text{ex}}=0.5$ for each trial. Further, we take the same cavity parameters as for the measurements presented above, i.e. a bare finesse of 17,500 (9,500) for the 580\,nm (611\,nm) transition, and a fiber radius of curvature of 25\,µm resulting in mode waists of 1.41\,µm (1.44\,µm) at the double-resonance condition at about 6\,µm cavity length, and 1.18\,µm for contact mode at 2.5\,µm cavity length. The cavity length jitter strongly affects the Purcell factor quantified by eq.\,(1) in \cite{pallmann_highly_2023} and thus the fluorescence count rate. Here, we set the cavity length stability to the best measured values of 2.5\,pm (0.8\,pm) for an open cavity (contact mode) as presented in \cite{pallmann_highly_2023}. The detected count rate can then be calculated as,
\begin{equation*}
    R_{\text{det}}=R_{\text{pulsed}}(d_{\text{NP}})\cdot \eta_{\text{out}}(d_{\text{NP}})\cdot T_{\text{path}}\cdot\eta_{\text{det}},
    \label{eq:R_det}
\end{equation*}
with the cavity outcoupling efficiency $\eta_{out}\leq0.55$, which is dependent on the nanoparticle scattering losses. The detector has an efficiency of $\eta_{\text{det}}\approx 0.65$ at a dark count rate of 20\,Hz and the transmission of the collection optics path amounts to $T_{\text{path}}=0.8$. All together, this results in a probability of up to 30\% to detect a photon which is emitted into the cavity mode. The pulsed emission rate into the cavity mode is given by:
\begin{align*}
    R_{\text{pulsed}}(d_{\text{NP}})=&p_{\text{ex}}\cdot f_{\text{rep}}\cdot\frac{F_{\text{P}}^{\text{eff}}(d_{\text{NP}})}{F_{\text{P}}^{\text{eff}}(d_{\text{NP}})+1}\\
    &\cdot \left(1-e^{-\frac{F_{\text{P}}^{\text{eff}}(d_{\text{NP}})+1}{T_{1,\text{fs}}}t_{\text{det}}}\right).
    \label{eq:R_pulsed}
\end{align*}

\begin{figure}
\includegraphics[width=\columnwidth]{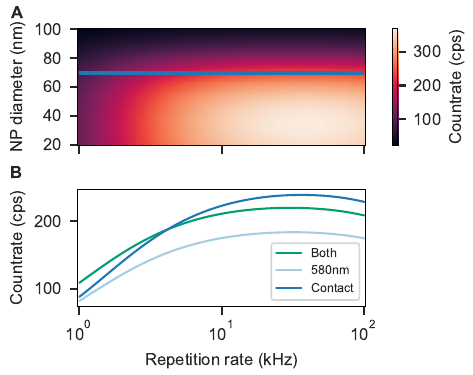}
\caption[Estimation of the single ion count rate for a pulsed resonant measurement.]{Calculation of the detected count rate for a perfectly coupling, single europium ion in a pulsed, resonant measurement scheme. \textbf{A}: the cavity is operated in contact mode with an RMS length jitter of 0.8\,pm. The blue line marks the nanoparticle size of 70\,nm assumed for the calculation in \textbf{B}. \textbf{B}: dependence of the count rate on the repetition rate for different operation modes of the cavity. At the double-resonance condition, the count rate can be increased by additionally collecting the 611\,nm fluorescence (both, green) compared to only the 580\,nm light (light blue).}
\label{fig:single-ion-countrate}
\end{figure}

The plot in Fig.\,\ref{fig:single-ion-countrate}A shows that in contact mode, where the cavity fiber touches the planar mirror and the smallest cavity length is achieved, count rates above 300\,cps could be reached for a 40\,nm NP, which should be within the size distribution of the batch used here. The plot also shows that smaller NPs lead to higher count rates since the Rayleigh-scattering losses are smaller. In this scenario, a maximum Purcell factor of 5.6 could be achieved. In Fig.\,\ref{fig:single-ion-countrate}B, a cut for a NP diameter of 70\,nm, the smallest measured so far, across the plot in A is shown together with the estimations for an open cavity configuration. Enhancing and collecting both transitions (green) can significantly increase the count rate compared to the case where only the 580\,nm transition is collected (light blue). However, for sufficiently high repetition rates, the contact mode provides even higher signals due to an increased cavity length stability and smaller mode waist.
Although the estimated count rates are still low compared to other solid state emitters, with the current setup, it should be feasible to detect single europium ions inside a 70\,nm NP in contact mode with a signal-to-noise ratio of about $SNR=240\,\text{cps}/\sqrt{20}\,\text{cps}=54$, according to Fig.\,\ref{fig:single-ion-countrate}B and the detector dark count rate of 20\,Hz.

\section{Conclusion}
We have performed cavity-enhanced spectroscopy of few-ion ensembles of europium inside individual yttria nanoparticles. Using an aerosol printer, we established a new method to disperse individual nanoparticles on the surface of the planar mirror in a controlled manner. Scanning cavity transmission and fluorescence scans were conducted to locate six different nanoparticles with diameters ranging from 70\,nm to 95\,nm. The widths of the inhomogeneous linewidths have been measured to be between 30\,GHz and 74\,GHz, which is broader than in bulk crystals but possibly advantageous for single ion selection. The high degree of control of the cavity length permits us to selectively enhance just the $^5D_0\rightarrow\,^7F_0$ transition, the $^5D_0\rightarrow\,^7F_2$ transition or both simultaneously. As expected, we observe the shortest lifetime when making use of the multimodal Purcell enhancement of both transitions and ideal (effective) Purcell factors up to 140 (1) have been measured. Furthermore, we were able to obtain an upper bound of the homogeneous linewidth of 3.3 (0.6)\,MHz of a sub-ensemble in the center of the inhomogeneous line by transient spectral hole burning. Based on these results, we estimated count rates of up to 300\,cps for an optimally-coupled single ion inside a 40\,nm NP. The measurement scheme that could be employed therefore is similar to the one used in \cite{deshmukh_detection_2023}, where they were able to detect the fluorescence from single erbium ions inside yttria nanoparticles in a similar fiber-based microcavity stage.

To summarize our findings, the complete set of parameters describing the cavity-emitter system for the 580\,nm as well as the 611\,nm transition, respectively, can be found in Tab.\,\ref{tab:cQED-parameters}. The figure of merit of such a system is the cooperativity $C$ defined as \cite{auffeves_controlling_2010}:
\begin{equation*}
    C = \frac{4g^2}{(\kappa + \Gamma_h)\,\Gamma_h},
    \label{eq:cooperativity}
\end{equation*}
which relates the cavity-emitter coupling rate $g$ to the loss rate of the cavity $\kappa$ and the dephasing rate of the emitter $\Gamma_h = 1/(2\pi\,T_1) + \Gamma_{\text{d}}$. Here, $\Gamma_d$ is the pure dephasing rate. 
\begin{table}[h]
    \begin{center}
    \begin{tabular}{l|c|c}
        Parameter & \textsuperscript{5}D\textsubscript{0}$\rightarrow$\,\textsuperscript{7}F\textsubscript{0} & \textsuperscript{5}D\textsubscript{0}$\rightarrow$\,\textsuperscript{7}F\textsubscript{2} \\
        \hline
        $\lambda$\,\textnormal{(nm)} & $580.8$ & $611$ \\
        $g\,$\textnormal{(MHz)} &  $2\pi\cdot 0.4$ &  $2\pi\cdot 2.4$ \\
        $\kappa\,$\textnormal{(GHz)} & $2\pi\cdot 1.8$ & $2\pi\cdot 2.5$ \\
        $\Gamma_{\textnormal{h}}\,$\textnormal{(MHz)} & $2\pi\cdot 3.3$ & $2\pi\cdot 0.7\cdot 10^6$ \\
        $F^{\textnormal{eff}}_{\textnormal{P}}$ & $3.4$ & $0.47$ \\
        $C$ & $8\cdot 10^{-5}$ & $5\cdot 10^{-11}$ \\
    \end{tabular}
    \end{center}
    \caption[Summary of cavity QED parameters]{Summary of the measured, best-case parameters describing the cavity-emitter system of a 70\,nm nanoparticle and $T_1=2.0\,$ms.}
    \label{tab:cQED-parameters}
\end{table}
For future quantum information applications it would be desirable to reach $C\approx 1$, which allows for a near-deterministic entanglement between the photonic and atomic quantum states as e.g. shown by \cite{daiss_quantum-logic_2021}. In order to reach this regime, improved optical coherence is crucial, e.g. $\Gamma_h = 2\pi \times 25$~kHz as has been observed for slightly larger Eu$^{3+}$:Y$_2$O$_3$ nanoparticles \cite{liu_controlled_2018}. With a small mirror radius of curvature ($r_c=8~$µm) and a short cavity length in contact mode, one can reach a mode waist of 0.7\,µm. Together with a finesse of 55,000, this reaches a unity cooperativity.
In this regime, efficient single-shot readout of hyperfine qubit states becomes possible, as well as entanglement generation between separate cavity-ion nodes \cite{reiserer_colloquium_2022}, laying the grounds for distributed quantum processing nodes.

\section{Author statements} 
\begin{acknowledgement}
We thank A. Quintilla and P. Brenner at the Nanostructure Service Laboratory (CFN-NSL) at KIT for carrying out the AFM and SEM measurements, and Julia Benedikter, Bernardo Casabone, and Thomas Hümmer for contributions in the early phase of the experiment. Further, we thank Chetan Deshmukh and Eduardo Beattie for many fruitful discussions. We acknowledge support from Leonhard Neuhaus to adapt the Pyrpl software package to our needs.
\end{acknowledgement}

\begin{funding}
This work has been financially supported by the European Union Quantum Flagship initiative under grant agreement No. 820391 (SQUARE), the Karlsruhe School of Optics and Photonics (KSOP), the BMBF project NEQSIS (contract no. 16KISQ029K), and the Deutsche Forschungsgemeinschaft (DFG) through the Collaborative Research Centre “4f for Future” (CRC 1573 project number 471424360, project C2). S. Liu acknowledges support by the National Natural Science Foundation of China (Grant No. 12304454), and Guangdong Basic and Applied Basic Research Foundation (Grant No. 2021A1515110191).
\end{funding}

\begin{authorcontributions}
\textbf{Timon Eichhorn:} Conceptualization; Data curation; Formal analysis; Investigation; Methodology; Software; Validation; Visualization; Writing - original draft; Writing - review \& editing. \textbf{Nicholas Jobbitt:} Conceptualization; Data curation; Formal analysis; Investigation; Methodology; Writing - review \& editing. \textbf{Shuping Liu:} Sample fabrication and characterization; Writing - review \& editing. \textbf{Diana Serrano:} Sample fabrication and characterization; Writing - review \& editing. \textbf{Robert Huber:} Aerosolprinting; Writing - review \& editing. \textbf{Tobias Krom:} Cavity fiber fabrication; Writing - review \& editing. \textbf{Sören Bieling:} Cavity fiber fabrication; Laser stabilization setup; Writing - review \& editing. \textbf{Ulrich Lemmer:} Funding acquisition;  Methodology; Project administration; Resources; Supervision; Validation; Writing - review \& editing. \textbf{Hugues de Riedmatten:} Conceptualization; Funding acquisition;  Methodology; Project administration; Resources; Supervision; Validation; Writing - review \& editing. \textbf{Philippe Goldner:} Conceptualization; Funding acquisition;  Methodology; Project administration; Resources; Supervision; Validation; Writing - review \& editing. \textbf{David Hunger:} Conceptualization; Funding acquisition;  Methodology; Project administration; Resources; Supervision; Validation; Writing - review \& editing.\\
\newline
All authors have accepted responsibility for the entire content of this manuscript and approved its submission.
\end{authorcontributions}

\begin{conflictofinterest}
Authors state no conflict of interest.
\end{conflictofinterest}

\begin{informedconsent}
Informed consent was obtained from all individuals included in this study.
\end{informedconsent}

\begin{dataavailabilitystatement}
The datasets generated during and/or analyzed during the current study are available from the corresponding author on reasonable request.
\end{dataavailabilitystatement}

\bibliographystyle{ieeetr}
\bibliography{refEu-cavity-spec}

\end{document}